# Emitter near an arbitrary body: Purcell effect, optical theorem and the Wheeler-Feynman absorber

Murugesan Venkatapathi

*Supercomputer Education and Research Center, Indian Institute of Science, Bangalore – 560012.*
Corresponding author: murugesh@serc.iisc.ernet.in

**Abstract**

The altered spontaneous emission of an emitter near an arbitrary body can be elucidated using an energy balance of the electromagnetic field. From a classical point of view it is trivial to show that the field scattered back from any body should alter the emission of the source. But it is not at all apparent that the total radiative and non-radiative decay in an arbitrary body can add to the vacuum decay rate of the emitter (i.e.) an increase of emission that is just as much as the body absorbs and radiates in all directions. This gives us an opportunity to revisit two other elegant classical ideas of the past, the optical theorem and the Wheeler-Feynman absorber theory of radiation. It also provides us alternative perspectives of Purcell effect and generalizes many of its manifestations, both enhancement and inhibition of emission. When the optical density of states of a body or a material is difficult to resolve (in a complex geometry or a highly inhomogeneous volume) such a generalization offers new directions to solutions.

**Introduction**

That even the spontaneous emission from an emitter is a function of the surrounding medium was first proposed by Purcell [1]. His observation of increased nuclear magnetic transitions by radio frequency emission when coupled to a resonant cavity led him to the argument on the increase in the density of states. After it was realized that excited atoms coupled to cavities have different decay rates than in free space, the effect of a nearby surface on the decay rate of an emitter was elucidated with a series of significant experiments [2]. These results were explained by the effect of the reflected field back at the source, the symmetric green tensors [3, 4] and their quantized versions [5]. In these approaches the total energy absorbed or radiated by the surface was not accounted for explicitly. Lately, spontaneous emission is typically described as a function of the local density of propagating states the emitter can be coupled to [6, 7]; these states are defined by the allowed modes of the classical field in the surrounding media (except when non-classical states are manifest). This approach has found wider mention in the effects of photonic crystals on spontaneous emission of an emitter [8, 9].

Problems with various emitter-receiver geometries for possible single photon communication, quantum dots in a photonic crystal for material applications, and other manifestations of Purcell effect are represented using the coupling between the classical modes and appropriately quantized emitters [10-13]. The Purcell enhancement factors of a quantum dot near a regular shaped body like a small metallic cylinder or a sphere has hence been elucidated for plasmon based communication applications. In material applications, when one or a *few* emitters like quantum dots (very low volume fractions) are embedded, the propagating modes of the surrounding material are found and the emitter is quantized and coupled to this material as a perturbation [14]. Even these approaches become intractable when we cannot resolve the propagating states of the material. When we have many quantum dots and other scatterers rather randomly distributed within a material, the optical density of states is a strong function of the emitter and cannot be determined independent of the quantized emitter. This is true for materials like hybrid films that exhibit interesting optical properties but are harder to explain quantitatively [15]. While the non-linear effects like saturation of emitter and others may not be included satisfactorily using classical methods, the classical field is easier to evaluate for complex inhomogeneous geometries. The question then is whether a direct action based on the classical field solutions (computed using methods like Finite-Difference-Time-Domain models and Discrete Dipole approximations) can be quantized instead of resolving the optical density of states in the inhomogeneous volume. While the answer to this question may not be readily available, the objective of this work is to elucidate Purcell effect without the function of density of states; and as a consequence revisit two other ideas of the past, the optical theorem and the Wheeler-Feynman theory of radiation. First, the altered spontaneous emission of a source near an arbitrary body is shown using conservation of energy in the classical electromagnetic field. It is shown that the effect of the scattered field back on the source is a universal function of the total radiative and non-radiative decay in the body irrespective of its geometric or optical properties. Secondly, it is to present another action-at-distance mechanism that can elucidate Purcell effect equally well; the Wheeler-Feynman theory of radiation. To quote Wheeler and Feynman, "the conventional expression for force of radiative reaction represents a statistical average only". Though they offered those comments on the effect of thermodynamic fluctuations on their theory of radiation [16], it is equally true of spontaneous emission of an emitter.

**I. Optical theorem for a point source – a physical interpretation**

The conventional optical theorem relates reduction of the energy of a plane wave beam when a body is introduced into it, to an equivalent reduction in the area of the original beam. This area called as the total/extinction cross section of the body is related to the incident field as in equation (1).

$$\sigma_{ext} = \frac{4\pi}{k} \Im[\mathbf{E}_i^* \cdot \mathbf{F}(\mathbf{k} = \mathbf{k}_o)] \qquad \text{Eq. (1)}$$

where $\mathbf{F}, \mathbf{k}_o$ are the scattering amplitude of body and incident momentum.

It is valid irrespective of the geometric and optical properties of the body. The optical theorem can be linked to Strutt's (Lord Rayleigh) and Sellmeier's idea [17] that the index of refraction is related to the absorption coefficient of a body (ie) the total scattering by the constituent particles of a body has to be related to the extinction of the incident beam in the forward direction. This idea was extended to scattering of particles by Feenberg [18]. Wheeler introduced the Scattering matrix approach and its unitarity [19]. The physical interpretation of this was alluded to by Bohr a few years since [20] and it was also later independently derived by Heisenberg [21]. Wick, Lax and Chiff [22-24] extended it with the inclusion of inelastic processes and spin. But as far as electrodynamics and optics are concerned, the discovery of the optical theorem is credited to van de Hulst [25]; the term *optical theorem* has been well known ever since [26]. His physical interpretation of the conventional optical theorem for an incident plane wave is insightful. When the detector measures the intensity of radiation purely in the forward direction, the light removed from the beam then should be only a function of the rate of energy absorbed and scattered by the body (i.e) its total extinction cross section. In other words, the area of its 'shadow' multiplied by the incident intensity should be the total energy scattered and absorbed from the plane wave field. When the incident beam is not an ideal plane wave field, this conventional optical theorem does not hold true; as has been shown with Gaussian beams [27]. Point sources on the other hand need a new approach, and the optical theorem has indeed been extended to the point sources in the last decade [28]. The salient conclusion of this optical theorem for point sources is that the scattered field back at the source is a function of the total/extinction cross section of the body as in equation (2).

$$\sigma_{ext} = \frac{4\pi}{k} \Im[\mathbf{E}_s^* \cdot \mathbf{p}] \text{ where } \mathbf{p} \text{ is the dipole moment vector of source} \qquad \text{Eq. (2)}$$

While we may not repeat this derivation, an alternative deduction based on local conservation of energy is pursued here. Using energy balance allows us to deduce Purcell effect and correspondingly expand the conclusions of the optical theorem for a point source. The total field *outside* the body in the presence of a point source *P* (figure 1) can be decomposed into the incident field and a scattered field. These two fields defined here are solutions of Maxwell's equations that satisfy the boundary conditions (in conjunction with an internal field valid only inside the body). While the incident field has its source inside an imaginary surface $S_o$, the scattered field will have its source inside the imaginary closed surface $S_1$ (that encloses the body). The total rate of energy flow '*W*' across any surface outside the body can then be decomposed into the incident energy, scattered energy and the extinction energy as in equation (3). By this definition, the incident and scattered energies have only sources and no sinks, and any work done by the incident and scattered fields have to be balanced by the flow of extinction energy. It should be noted that these three energies are not *independently* measurable physical quantities, and as mentioned before, an incident plane wave from a source at infinity is a scenario where these energies can be reasonably resolved by measurements.

$$\mathbf{E}_t = \mathbf{E}_i + \mathbf{E}_s \ , \ \mathbf{H}_t = \mathbf{H}_i + \mathbf{H}_s \ , \ W = \frac{1}{2}\int \mathbf{n} \cdot \Re(\mathbf{E}_t \times \mathbf{H}_t^*)dA$$

$$W_s = \frac{1}{2}\int \mathbf{n} \cdot \Re(\mathbf{E}_s \times \mathbf{H}_s^*)dA \ , \ W_i = \frac{1}{2}\int \mathbf{n} \cdot \Re(\mathbf{E}_i \times \mathbf{H}_i^*)dA \ ,$$

$$W_{ext} = \frac{1}{2}\int \mathbf{n} \cdot \Re(\mathbf{E}_s \times \mathbf{H}_i^* + \mathbf{E}_i \times \mathbf{H}_s^*)dA \ \Rightarrow W = \pm W_i \pm W_{ext} \pm W_s$$

Eq. (3)

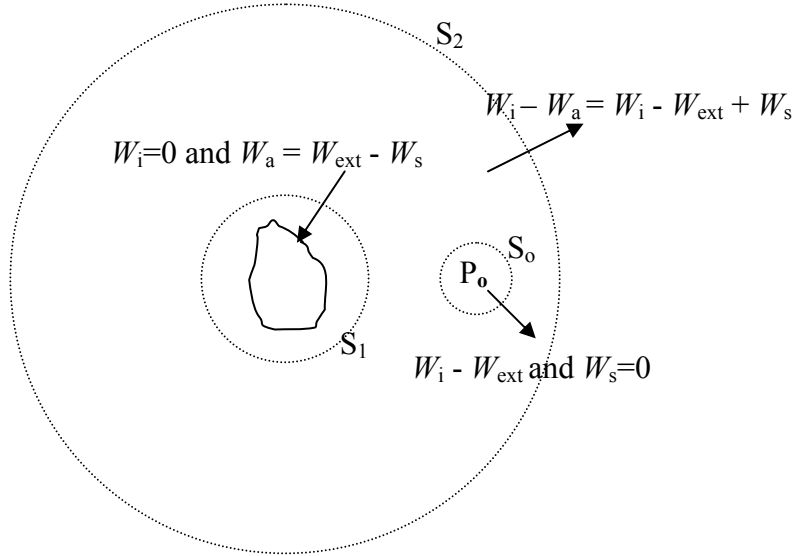

Figure 1: Energy balance around the body in the presence of a point source. Note that $S_o$ is an *imaginary* closed surface (like $S_1$ and $S_2$) of infinitesimal area around the source.

These $W$'s are >0 or <0 depending on the choice of the inward or outward normal (***n***) to the surface of integration; the direction of energy flow and their signs can be trivially assigned by physical deduction. Using an energy balance (time averaged) across an imaginary closed surface $S_1$ that contains only the body in a non-absorbing medium, it is evident that the energy absorbed by the body '$W_a$' into this closed surface is $W_{ext} - W_s$, as the net $W_i$ flowing through this surface is zero (scattered energy is evidently going out of the enclosed volume). Similarly, an energy balance across the imaginary closed surface

S$_2$ containing both the source and body in a non-absorbing medium should give an energy outflow of $W_i - W_a$. Now, an energy balance across the imaginary surface S$_o$ that encloses only the source should have net $W_s=0$ leaving behind $W_i - W_{ext}$. But an outflow of energy $W_i$ across S$_o$ is required for the overall energy balance across S$_1$ and S$_2$ to hold as well, by definition. This means the additional work done by the source due to the scattered field has to result in extinction energy $W_{ext}$ through S$_o$. Symmetry demands that the extinction energy ($W_{ext}$) through the surfaces S$_1$ and S$_o$ are equal as they are generated by the same set of source fields (see Appendix). So firstly, this energy balance demands that the extinction cross section is the total cross section (ie) sum of absorption and scattering cross sections ($W_{ext} = W_a+W_s$) even for point source excitation of a body (as in the case of plane wave excitation). It is obvious that in the presence of the body, its scattered field back on the source (coupling with the source) will have to change its emission. Here we find that this additional work done by the source has to be proportional to rate of extinction (ie) sum of the rates of energy absorbed and scattered by the body in all directions. The change in work done by the source (emission) is just its quality of resonance with the field scattered back (i.e.) $\Im[\mathbf{E}_s^* \cdot \mathbf{p}]$. The conjugation of the scattered field reflects the additional force the source has to work 'against' when its amplitude $p$ is fixed, unlike a driven oscillator at resonance where the amplitude increases. This extinction energy $W_{ext}$ is the rate of flow of momentum through an effective area of the body called the extinction cross section of the body as in equation (2).

This derivation is valid for even non-harmonic fields and a body in the near-field of the source. But the optical theorem is not concerned with the additional work done by the source in the presence of the body. It rather confined itself to the fact that the total cross section of the body is a universal function of its scattered field back at the point dipole source. The above interpretation of the optical theorem means that a change in decay rate of an emitter in proportion to the extinction of the body is *expected*. It should be also emphasized that this interpretation does not merely relate change in emission of the source to the energy scattered back at it, which is relatively trivial. The significant observation is that this *change* in emission is a function of the total energy scattered and absorbed by a body. This establishes that the total decay rate of the emitter is the sum of its vacuum decay rate, the radiative decay rate (scattering) and the non-radiative decay rate (absorption) due to the body. For metallic bodies smaller than wavelength the non-radiative decay rate of the body defined here includes the decay into plasmons. It should be noted that the decay of the emitter into the guided modes of an object like an infinite cylinder placed near it is included in this definition of extinction as well. When an emitter is placed very close to the body, the radiative and non-radiative decay rates into the body ($W_{ext}$) can be much larger than its vacuum decay rate ($W_i$-$W_{ext}$); such high decay rates into a body are relevant for applications like single photon switching [12]. It is also useful that for bodies smaller than a few wavelengths in dimension the scattering and extinction cross sections can be significantly larger than their geometric cross sections [29]. For small metallic particles under plasmon resonance, this extinction cross section can even be a few orders of magnitude larger than their geometric cross sections.

If the point source is a harmonically oscillating dipole moment, this additional work done on the oscillation is well inferred as faster decay rate. For a typical emission life-time of $10^{-8}$ sec in the visible range of spectrum (having $10^7$ oscillations), the continuous harmonic oscillator model of the source is a good approximation. That is more than sufficient time for energy to be absorbed by a nearby body (less than few hundred wavelengths away) and coupled back to the source as represented in a classical model. The relative decay rate $\gamma$ for a point dipole source with an amplitude $p$ (with non-relativistic motion of the charges) is then given by

$$\frac{\gamma}{\gamma_o} = 1 + \frac{\Im[\mathbf{E}_s^* \cdot \mathbf{p}]}{\frac{2p^2\omega^3}{3c^3}} = 1 + \frac{3}{2}\frac{\Im[\mathbf{E}_s^* \cdot \mathbf{p}]}{p^2 k^3} \qquad \text{Eq. (4)}$$

Using the above relation, the change in rate of spontaneous emission near a body can be found by calculating the field scattered back at the source. The scattered field can be calculated using classical electrodynamics when the geometry and the optical properties of the body are known. The computational formulations like DDA and FDTD that use spatially discretized representations are convenient in including excitation from point sources. On the other hand, the coupling of the radiative and non-radiative decay in the body as a function of its size and distance from source can be derived using analytical mode decomposition [11, 12]. A significant point to note is that the change in spontaneous emission by increasing the size of a body is not unbounded. The extinction cross section can increase with both the thickness and the geometric cross section of the body without limit. But the 'light time' to all parts of the body has to be much smaller than the lifetime of the emitter to couple the energy back to it. These conclusions are reflected even in the early and the more recent experiments with emitters near surfaces; the enhancement of spontaneous emission becomes negligible at large distances compared to wavelength regardless of the large extinction cross section of the surface [2, 31]. Also, since a typical emission is not single harmonic and consists of a band of frequencies, the presence of the body can result in a small change in its mean frequency of emission. The extinction cross section of a body can change significantly with the frequency especially near its absorption bands; thereby the highly frequency dependent back-scattered field of the body alters the spectrum of the source correspondingly. This shift can be towards or away from the resonance of the body depending on whether the body enhances or inhibits the emission.

It is evident that the emission from the source can be inhibited by the body if the phase of the scattered field back at the source is around $\pi/2$ radians ahead of the oscillation and also significant in its amplitude (representing absorption at source). This means a *negative* extinction cross section as per equation (2), but as unphysical as it appears, it just means here that the source is inhibited in proportion to the radiative and non-radiative decay in the body. It is also obvious from equation (4) that the polarization of the scattered field influences the emission that can result in non-isotropic emission under the influence of the body. These non-isotropic and inhibited emissions are observed in the band gap frequencies of photonic crystal and in cavities. Thus this energy balance in the classical electromagnetic field clearly *mandates* the Purcell effect; a change (enhancement or inhibition) in the spontaneous emission of a source near a body in proportion to its extinction cross section. Nevertheless, the fact that an emission can get scattered from a body which in turn affects the source moments later may not seem completely satisfactory as a mechanism for the change in spontaneous emission. A similar conundrum had challenged Wheeler and Feynman while framing their theory of radiation resistance that occurs by action of the surrounding matter, but simultaneous with the emission. Their theory of radiation which is a precursor to Purcell's discovery of change in spontaneous emission, predicts such an effect when applied to a source placed near a body.

**II. A small body as a partial Wheeler-Feynman absorber**

The Wheeler-Feynman (WF) absorber theory of radiation was formulated to explain away the self interaction of an electron. But to accommodate phenomenon like the Lamb shift discovered later, the self-interaction term was found necessary and hence this theory was of no further use there. Nevertheless, this theory provides a clear elucidation of the classical force of radiative reaction on an accelerating charge. This theory has thus been used to elucidate other problems like the arrow of time [32], expansion of universe [33], and scattering of particles [34]. The significance of the WF absorber is the use of time-symmetrical solution of Maxwell's equations as opposed to only the retarded solution; Maxwell's equations are invariant under a time inversion. The essential idea of this theory of radiation was first proposed by Tetrode; that an emitter can emit only in the presence of other matter in the universe. It was interpreted by Wheeler and Feynman that the source of radiation reaction comes from the matter in the universe and hence no self-interaction required for an accelerating charged particle to emit radiation.

They proposed a mechanism for the origin of radiation reaction purely from the matter exclusive of the source [16, 35]; this entailed half-advanced and half-retarded emission from any emitter. They assumed that an *isolated* charge emits half-advanced and half-retarded waves such that there is no net emission and hence no self-interaction. This is true if the advanced wave emitted is just the retarded wave of the emitter with an inversion in time, and consequently momentum as well (the other advanced solution involves both time and space inversion). The retarded waves then have outgoing momentum while advanced waves have momentum incoming to the source. This implies no net change of momentum, or a combination of emission and absorption that exactly cancel each other and it is an equally valid solution of the Maxwell's equations. In the presence of matter though, the retarded waves from the source are absorbed by the charges around which in turn emit such half-advanced and half-retarded waves as well. While the retarded waves from the source and advanced waves from the matter constructively add to the full retarded waves we measure, the advanced waves from the matter cancel advanced waves from the source leaving no measurable advanced effects 'except at the source'. In other words, the absorber (matter around emitter) increases the *apparent* difference between the retarded and advanced waves giving rise to a net emission. The emission is then purely a result of the material around the accelerating charge. The effect of the advanced waves from the matter around the emitter was shown to be a force equal to the radiative reaction at the source, right at the instant of initial acceleration of the charge. Since this force is in fact the third derivative of the position of the charge, a small pre-acceleration of the charge before the instant of emission was unavoidable in this theory; nevertheless this was shown to be too small to be detected by experiments. This theory was valid as long as the matter around the source was a complete absorber. A complete absorber is a material that has at least a thickness *$2\lambda/(n-1)$* all around the source, where *n* is the refractive index of the material; a condition the Universe certainly satisfies. This also implies there are no retarded or advanced waves that propagate beyond the absorber, and the total radiation reaction of the source and the absorber sum to zero. The shape of the complete absorber or its thickness beyond this limit has no effect on the force of radiative reaction on the source.

If we have a source *P* that emits half-advanced and half-retarded waves as in the WF radiation theory (equation (5) and figure 2), all particles in the complete WF absorber receive the retarded waves of the source and in turn emit such half-retarded half-advanced waves. The sum of the advanced waves from all the charges in the absorber can be shown to add to a field apparently originating from infinity and converging on the source right at the beginning of emission. This field was proposed by Dirac earlier to account for the force of radiative reaction as shown in equation (5).

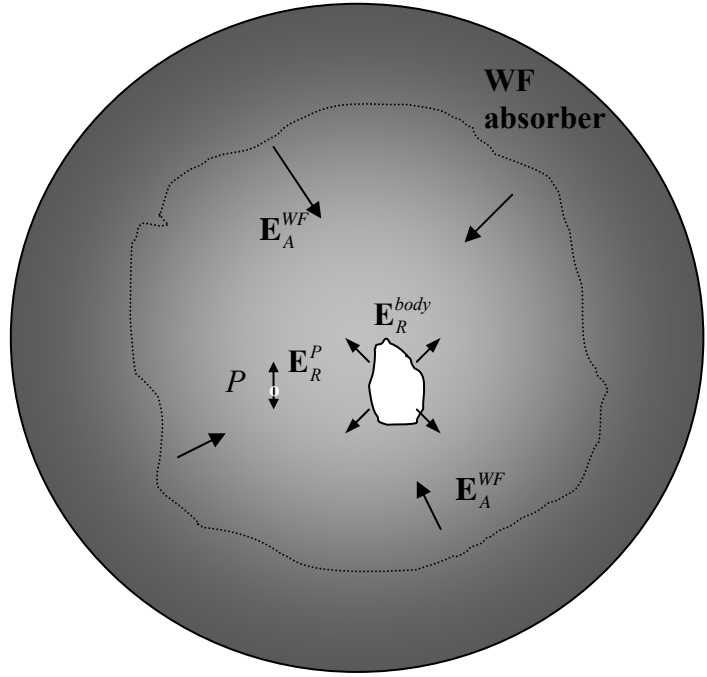

Figure 2: The electric components of fields acting on the source in the presence of the body and the WF absorber. The shape of the absorber and the body do not change the results or conclusions of the above and are for illustrative purposes only.

Dirac's radiation field $= [\mathbf{E}_R^P - \mathbf{E}_A^P]/2 = \mathbf{F}_o$ (reaction force on unit charge at source)

Source field $= [\mathbf{E}_A^P + \mathbf{E}_R^P]/2$

Dirac's field + Source field = full retarded field from source ($\mathbf{E}_R^P$)   Eq. (5)

$\sum_{all\ charges} [\mathbf{E}_R - \mathbf{E}_A]/2 = 0$ (total reaction) $\Rightarrow \mathbf{F}_o$ (at source) $= - \sum_{absorber} [\mathbf{E}_R^{WF} - \mathbf{E}_A^{WF}]/2$

If we add a small body very close to the source while rest of the matter is in the far-field acting as a complete WF absorber, the resulting radiation reaction is a first order perturbation of the system (figure 2). This is valid as long as the dimensions of the body are less than λ and by inference its volume much smaller than that of the complete WF absorber participating in this emission. This corresponds exactly to the condition proposed by Purcell in using small metal particles for the increased nuclear magnetic transitions by radio frequency emission [1]. The advanced effects from the complete WF absorber in this case can be decomposed into a complement of the source field and a complement of the scattered field from the body. As shown in the derivation of WF absorber theory [16], the advanced effects of a complete absorber due to the source field can manifest only at the source. Similarly the advanced waves from the absorber due to the scattered field act only at the charges of the body as the radiation reaction. But the advanced field from the body can indeed act on the source; the body is a partial WF absorber that is small enough to be a first order perturbation. This field from the body results in a modified force of radiative reaction. When the body is large or is far away from the source as the other matter is, it becomes a part of the complete absorber and hence no change in radiation reaction due to it. Mathematically, the above can be described by the modified radiation reaction on the source, when the condition of complete absorption is imposed.

$$\text{Total reaction} = \sum_{absorber} \frac{1}{2}[\mathbf{E}_R^{WF} - \mathbf{E}_A^{WF}] + \sum_{body} \frac{1}{2}[\mathbf{E}_R^{body} - \mathbf{E}_A^{body}] + \frac{1}{2}[\mathbf{E}_R^P - \mathbf{E}_A^P] = 0$$

$$\Rightarrow \mathbf{F}(\text{at source}) = -\sum_{absorber} \frac{1}{2}[\mathbf{E}_R^{WF} - \mathbf{E}_A^{WF}] - \sum_{body} \frac{1}{2}[\mathbf{E}_R^{body} - \mathbf{E}_A^{body}])$$

Eq. (6)

Now the crucial observation is that there are two extreme conditions associated with this expression in equation (6). The radiation reaction of the body (results in energy absorbed and scattered by the body) can exactly cancel the contribution from the absorber which implies inhibited or zero emission and it can happen under specific conditions as in the band gap of a photonic crystal or inside a cavity. A non-isotropic radiation reaction force, for example inside a 2D photonic crystal also manifests under this model and can thus result in preferential emission in a direction depending on the absorber or the body. But the more typical observation is when the radiation reactions from the absorber and the body sum resulting in enhanced emission. The advanced field of the body at the source is just the retarded scattered field of the body time inverted, hence $\mathbf{E}_s^*$. The part of this additional reaction force in phase with the third derivative of oscillation of the source gives the change in the force of radiation reaction.

$$F = F_o + \frac{\Re\{\frac{1}{2}[\mathbf{E}_s - \mathbf{E}_s^*] \cdot -i\mathbf{p}\}}{p} = F_o + \frac{\Im[\mathbf{E}_s^* \cdot \mathbf{p}]}{p}$$

Eq. (7)

With the above radiation reaction force due to the presence of the body, the emission from the source is again altered as in equation (4). This change of decay rate of an emitter in the presence of the body as given by Wheeler-Feynman theory of radiation is valid for a body of a general shape and optical properties. A significant question that arises in the WF theory is if the interaction of the charges in the body with each other is included. It should be noted that it is this interaction of retarded waves within the body (as is the case with absorber) manifests as the refractive index of the body and thus determines the time of action of each constituent charge and also the net scattered field from the body.

**Conclusion**

The optical theorem for a point source was deduced by a balance of energy in the incident, scattered and extinction energy in the classical electromagnetic field. This allowed us to generalize the altered emission of a source near a body of arbitrary shape and properties. It was shown that the decay rate of the point source in the presence of any body includes its vacuum decay rate and the radiative and non-radiative decay rates due to the body; a significant conclusion from a purely classical deduction. The Wheeler-Feynman theory of radiation is also a suitable candidate to classically elucidate Purcell effect because of its premise that the forces of radiative reaction come from the matter around an emitter. There the radiation reaction on all the charges in the body was explicitly shown to add constructively or destructively to the radiation reaction on the emitter. This leads to the same conclusions that the addition of the body's radiative and non-radiative (absorption) channels can result in increased emission from the source. Additionally the Wheeler-Feynman absorber theory explicitly requires a small body close to emitter for the above conclusions to be valid based on a first order perturbation.

**Appendix: Symmetry of extinction by two sources**

It is shown that the flow of the extinction energy of any source and an arbitrary scattering body is equal over any two imaginary closed surfaces around them. In the context of the work here, it means that the flow of extinction energy through surface $S_o$ is equal in magnitude to the extinction energy through $S_1$ (figure 1). The direction of flow through the surfaces could be independent, where one flows in and other flows out of the closed surface, or both flow in/out of closed surfaces, meaning odd and even symmetry. Specifically, we assumed there, that we have a point source that has net emission and a body that can only have a net absorption, which implies extinction energy always flows into the surface $S_1$ while it can flow in or out of $S_o$ (implying enhancement or inhibition of emission). There are other physical arguments that demand that the extinction of two sources by each other be symmetrical (odd or even) and we will not pursue them here, as also the application of vector Green's theorem. Instead we do an explicit integration of extinction of two generalized sources.

Let $\sum_{m,n} \mathbf{F}^i_{mn} g_n(\|\mathbf{r}'-\mathbf{r}_1\|)$ and $\sum_{m,n} \mathbf{F}^s_{mn} g_n(\|\mathbf{r}'-\mathbf{r}_o\|)$ be vector fields in 3 dimensions that represent two sources (of incident and scattered fields) centered at positions $r_1$ and $r_o$ inside imaginary closed surfaces $S_1$ and $S_o$ respectively. The sources are assumed to be harmonic in time which means the vectors $\mathbf{F}g$ are solutions of the vector Helmholtz equation $\nabla^2 \mathbf{F}g + k^2 \mathbf{F}g = 0$ (except at points $r_1$ and $r_o$ where they originate). Here we assume $\mathbf{F}$ is purely a function of two independent angles θ, φ and $\mathbf{F}g$ can represent the electric or magnetic components of the field ($\mathbf{E}(\mathbf{r}'), \mathbf{B}(\mathbf{r}')$). In spherical coordinates this

implies $g_n$ is a function $f(\frac{1}{r}, z_n, \frac{\partial z_n}{\partial r})$, $z_n$ being the spherical Bessel functions. Such a generalized representation is valid for realizable sources, whether a spherical point source or a scattered field from a body of complex geometry. What is not known is the existence of a one-to-one mapping between these two sets of vectors $\mathbf{F}_{mn}$, and hence if $\mathbf{F}_{mn}^s$ can be obtained from $\mathbf{F}_{mn}^i$ independently for each $m,n$, using the boundary conditions and the constitutive relations of electric and magnetic components. Such a one-to-one mapping exists only for a few types of sources and regular geometries of body (that are completely defined by a set of separable coordinates of Helmholtz operator), and for a limited set of other geometries an all-to-one mapping that converges in finite $m,n$ is possible (T-matrix methods). But the existence of such solutions for $\mathbf{F}_{mn}^s$ is not the subject of our pursuit here, and we remove all limitations on the geometry of the scattering body by assuming $\mathbf{F}_{mn}^s$ and $\mathbf{F}_{mn}^i$ are independent (meaning this treatment is valid for even two sources with net emission). Our objective here is to only show that the extinction of two such source fields integrated over the two surfaces ($I_1$ and $I_2$) is equal. This is described by

$$\left| \int_{S_o} \sum_n [g_n(\|\mathbf{r}'-\mathbf{r}_o\|) g_n(\|\mathbf{r}'-\mathbf{r}_1\|)] \sum_m (\mathbf{F}_{mn}^s \times \mathbf{F}_{mn}^i) \cdot \mathbf{n} \, dA_o \right|$$

$$- \left| \int_{S_1} \sum_n [g_n(\|\mathbf{r}'-\mathbf{r}_o\|) g_n(\|\mathbf{r}'-\mathbf{r}_1\|)] \sum_m (\mathbf{F}_{mn}^s \times \mathbf{F}_{mn}^i) \cdot \mathbf{n} \, dA_1 \right| \qquad \text{Eq. (A.1)}$$

$$= |I_1| - |I_2|$$

Taking the summation over $n$ outside the integral results in the residue for any $n$, and *sufficient* but *not necessary* conditions on the integrand can be derived for all $\theta, \varphi$.

if $g_n(\|\mathbf{r}'-\mathbf{r}_o\|) g_n(\|\mathbf{r}'-\mathbf{r}_1\|)\big|_{\mathbf{r}' \in S_o} = \pm g_n(\|\mathbf{r}'-\mathbf{r}_o\|) g_n(\|\mathbf{r}'-\mathbf{r}_1\|)\big|_{\mathbf{r}' \in S_1}$ (a)

and $[\sum_m (\mathbf{F}_{mn}^s \times \mathbf{F}_{mn}^i) \cdot \mathbf{n}] dA_0 = \pm [\sum_m (\mathbf{F}_{mn}^s \times \mathbf{F}_{mn}^i) \cdot \mathbf{n}] dA_1$ (b) Eq. (A.2)

then $|I_1| - |I_2| = 0$

Both these conditions are readily enforced for arbitrary shaped closed imaginary surfaces $S_1$ and $S_o$ centered at $r_1$ and $r_o$, when both surfaces are identical in geometry. This results from enforcing the positive sign of condition (a) and one of the two possibilities of condition (b), *identically* for all $n, m$. The choice of normal (inwards or outwards) and condition (b) together determine the odd or even symmetry of energy flow through the surfaces. This assumption of geometrically identical surfaces does not result in a loss of generality, as this case is physically no different compared to when the two imaginary surfaces are not geometrically identical. A closer look at these sufficient conditions is encouraging (equation A.2); for if the conditions (a) and (b) are to be satisfied uniformly over all $\theta$ and $\varphi$, such that sign of the integrand is fixed, not one, but two of the four possible combinations in signs are valid. Also this choice of sign conservation has to be met only *independently* for each $n, m$, which means the rigid restriction of the identical geometries is really not necessary. In showing the validity of our claim, we have used the validity for the most explicit of the sufficient criteria (ie) arbitrary but identical geometries of the imaginary closed surfaces, where all the signs of condition (a) and (b) are identically fixed for all $\theta, \varphi, n, m$, but this is not a necessary condition at all.